\documentclass[aip,amsmath,amssymb,preprint]{revtex4-1}

\usepackage{graphicx,amssymb,bm,amsfonts,amsmath,graphics,color,epstopdf}
\usepackage[bookmarks=false,pdfstartview=FitH,colorlinks=true,citecolor=blue,linkcolor=blue]{hyperref}
\usepackage{lineno}
\usepackage{physics}
\begin{document}

\title{Signature of multilayer graphene strain-controlled domain walls in quantum Hall effect}

\author{Paul Anderson}
\thanks{PA and YH contributed equally to this work.}
\affiliation{Department of Physics and Astronomy, California State University Long Beach, Long Beach, California 90840, USA}

\author{Yifan Huang}
\thanks{PA and YH contributed equally to this work.}
\affiliation{Department of Mechanical and Materials Engineering, University of Nebraska, Lincoln 68588-0526, Nebraska, USA}

\author{Yuanjun Fan}
\affiliation{Department of Mechanical and Materials Engineering, University of Nebraska, Lincoln 68588-0526, Nebraska, USA}

\author{Sara Qubbaj }
\affiliation{Department of Physics and Astronomy, California State University Long Beach, Long Beach, California 90840, USA}

\author{Sinisa Coh}
\affiliation{Department of Mechanical Engineering and Materials Science and Engineering Program, University of California, Riverside, Riverside, CA, 92521 USA}

\author{Qin Zhou}
\affiliation{Department of Mechanical and
Materials Engineering, University of Nebraska, Lincoln 68588-0526, Nebraska, USA}

\author{Claudia Ojeda-Aristizabal}
\affiliation{Department of Physics and Astronomy, California State University Long Beach, Long Beach, California 90840, USA}

\date{\today}

\begin{abstract}
  Domain walls, topological defects that define the frontier between regions of different stacking in multilayer graphene, have proved to host exciting physics. The ability of tuning these topological defects in-situ in an electronic transport experiment brings a wealth of possibilities in terms of fundamental understanding of domain walls as well as for electronic applications. Here, we demonstrate through a MEMS (micro-electromechanical system) actuator and magnetoresistance measurements the effect of domain walls in multilayer graphene quantum Hall effect. Reversible and controlled uniaxial strain triggers these topological defects, manifested as new quantum Hall effect plateaus as well as a discrete and reversible modulation of the current across the device. Our findings are supported by theoretical calculations and constitute the first indication of the in-situ tuning of topological defects in multilayer graphene probed through electronic transport, opening the way to the use of reversible topological defects in electronic applications.       
\end{abstract}

\pacs{}

\maketitle

Topological defects are behind intriguing phenomena in multilayer graphene. Examples are partial dislocations also known as domain walls, found at the borderline between different stacking orders. Domain walls occur naturally in multilayer graphene, and are restricted to the basal plane which imposes an extreme boundary condition that results into confinement \cite{Amelinckx, Butz}. Bilayer graphene, being the thinnest material that can host such topological defects has revealed through transmission electron microscopy (TEM) the dynamics and patterns of these defects, that are known to behave as solitons \cite{McEuen}. It has been found that electronic transport along the defects occurs through valley-polarized chiral electrons \cite{Long, Martin, Zhang, Vaezi}. Additionally, it is known that some configurations of partial dislocations completely block electronic transport across the dislocations while others don't have an effect, providing a possible explanation for the both observed metallic and insulating behavior in bilayer graphene at the Dirac point \cite{San-Jose, Shallcross}. Despite the fact that domain walls play an important role in the electronic properties of multilayer graphene, their effect on quantum Hall effect experiments has been so far unknown. 

In general, identifying the effect of topological defects in a conductance measurement is challenging. Electronic devices are usually not compatible with imaging techniques such as TEM. While it has been demonstrated that topological defects can be identified through near-field infrared nanoscopy \cite{Long}, simultaneous tuning of the defects and conductance measurements remain out of reach. Additionally, being domain walls naturally occurring in multilayer graphene, their contribution to electronic transport is usually buried within other effects such as charged impurities and other defects. Here we report the experimental signature of domain walls in the quantum Hall effect (QHE) measured in a multilayer graphene sample. The domain walls are dynamically created by uniaxial strain applied to the sample using a MEMS (micro-electromechanical system) actuator, while simultaneously measuring electronic transport at low temperature with an external magnetic field. Theoretical calculations are presented to support our findings.

With the emergence of the field of 2D materials, strain has provided an external tunable parameter that leads to exciting physics and applications, from strain-induced gauge fields that result into enormous pseudo-magnetic fields in graphene \cite{Guinea, Crommie} to devices capable of detecting fine local deformations \cite{Sakhaee-Pour, Kumar, Wang}. A few mechanisms have been proved successful in reversibly tuning strain in layered materials\cite{PerezGarza, Nadya}, leading to measurable effects despite layered materials' typical large stiffness \cite{Hone, Bertolazzi}. Here we have used a MEMS actuator to controllably create uniaxial strain in suspended multilayer graphene in a reversible and controlled manner, modifying the domain wall landscape in multilayer graphene. The emergence of domain walls upon stretching of multi-layer graphene sheets have been predicted in previous works, where the domain walls are consequence of the tension not being distributed evenly across all the graphene layers \cite{KumarDong, Yang}.

We have observed a strain-induced effect on QHE features as well as a discrete modulation of the current across the device, that we attribute to a strain-induced tuning of topological line defects in the multilayer graphene. Despite the fact that strain in graphene has been associated to the formation of large pseudo-magnetic fields \cite{Guinea, Crommie}, here the strain being uniaxial, triggers gauge fields with opposite signs at each valley, creating a zero-averaged pseudo-magnetic field in the sample. \\
In order to understand our experimental results, we have simulated numerically the effect of a domain wall in bilayer graphene QHE, finding additional QHE plateaus that are absent if no domain walls are included. In the following, we present the outcome of these calculations, that consider a 1D domain wall on the plane of a bilayer graphene, perpendicular to the electron flow, and subject to a magnetic field perpendicular to the plane of the sample, before turning to the experimental results.   

\begin{figure}
\includegraphics[width=15cm]{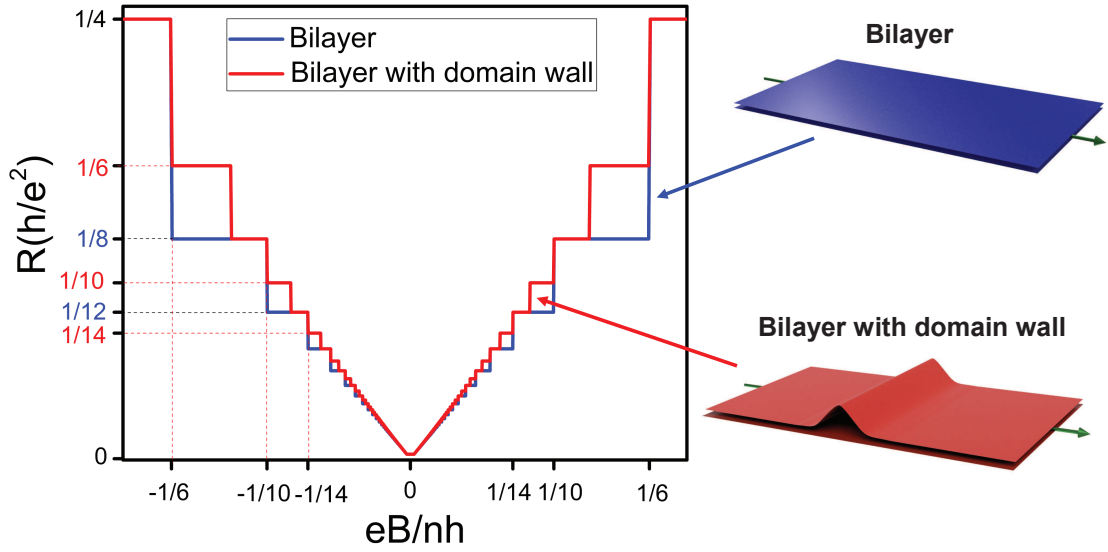}
\caption{\label{Sinisa1}a) Hall resistivity for a  bilayer graphene contrasted with the effect of adding a domain wall orthogonal to the electron transport. Bilayer QHE plateaus are indicated in black. Additional plateaus appear at $(1/6)(h/e^2)$, $(1/10)(h/e^2)$, $(1/14)(h/e^2),\ldots$, marked in red}
\end{figure}

In general, bilayer graphene is known to present an unconventional QHE, consequence of the coupling of the two graphene layers that turns graphene's Dirac Fermions into chiral quasiparticles with a quadratic dispersion relation, resulting in a characteristic Landau quantization. Figure \ref{Sinisa1} shows a schematics of the conventional integer QHE typical of a bilayer graphene. The transverse resistivity shows a plateau everytime there is a filled integer number of Landau levels, determined by the filling factor $\nu=nh/eB$, with $n$ the electronic density and $h$ Planck's constant. In the bilayer graphene this corresponds to $\nu=nh/eB=4N$ with $N=1,2,3,\ldots$. The transverse resistivity takes the form $\rho=(1/4N)(h/e^2)$ with $N=1,2,3,\ldots$ corresponding to $\rho=(1/4)(h/e^2)$, $(1/8)(h/e^2)$, $(1/12)(h/e^2)$, etc. Interestingly, when a domain wall is added perpendicular to the electron flow, new plateaus appear at $\rho=(1/6)(h/e^2)$, $(1/10)(h/e^2)$, $(1/14)(h/e^2),\ldots$  as shown in Figure \ref{Sinisa1} and in consistency with a recent work \cite{Babler}. 

It has been demonstrated in the past that QHE in twisted bilayer graphene presents the same plateau resistance values as a commensurate Bernal bilayer, consequence of an absence of symmetry breaking \cite{Lee, Mucha}. While a domain wall has a similar effect on the stacking of the layers as found in a twisted bilayer graphene, we have found that the former has a clear effect on the quantization of the resistivity. Our computed density of states for the bilayer graphene in the presence of a domain wall shows however no splitting of the Landau levels. 

We now turn to the experimental results, where device fabrication played a key role. We started with our unique MEMS chips, for which the top surfaces are polished to extremely flat faces with only a few nanometers of surface topology variation, with no deep trenches or holes in contrast with other methods \cite{Bernal}. Such a flat surface allows an easy transferring, patterning and electrical contacting of the multilayer graphene by nanofabrication techniques. 
Transfer microscopes and micromanipulators allowed to locate the multilayer graphene and transfer it into precise locations of the MEMS chips. This process was followed by nanolithography to pattern and define clampling metal electrodes and finally, the silicon dioxide sacrificial layer was etched away followed by critical point drying. The integrated graphene was suspended with both ends anchored by metal supports onto a movable structure supported by flexural beams. The metal supports anchor the graphene sample from its top surface, creating the asymmetry in tension force needed for the creation of domain walls \cite{KumarDong, Yang}. The tension forces are generated by the connected MEMS actuators which utilize electrostatic force. A scanning electron microscopy (SEM) image of the sample is shown in Figure \ref{Sample}. Detailed fabrication process is described in the Supplementary Materials.

Thanks to the fact that MEMS are compatible with high vacuum and cryogenic temperatures, we were able to probe the strain-induced effects through electronic transport in a closed cycle cryostat with a superconducting magnet. Electronic transport measurements were performed in a temperature range of 1.5 K - 300 K and magnetic fields up to 12 T. Differential conductance was measured by superimposing an AC voltage ($800\ \mu$V) to a DC bias voltage (up to $90$ mV) and measuring the current modulation by lockin detection. Similarly, differential resistance at zero bias was measured imposing an AC current ($10$ nA) and measuring the DC voltage drop across the sample. 

Despite the fact that domain walls are present in multilayer graphene and do influence electronic transport, their contribution is difficult to discriminate. It is only through the effect of strain and continuous electronic transport measurements that the impact of such defects becomes discernible. Strain was introduced by the electrostatic actuator, where a DC voltage was applied to the MEMS actuators in the range $0\ V$ - $80\ V$ to control the tension on the multilayer graphene. This voltage range imposes an averaged strain level of $0-0.21\%$ in the graphene sample, as estimated through the hybrid device model detailed in the Supplementary Materials. Our finite element analysis revealed a non-uniform strain distribution where the strain at the corners of the sample is the highest, where we believe the domain walls are triggered before traversing the sample. We observed that the effect of the strain was reversible, reflected both on the sample resistance and on the features of the differential conductance and magnetoresistance. 

Magnetoresistance measurements are presented in Figure \ref{R_B_V-2}. We fixed the actuation voltage and continuously varied the magnetic field as we  measured the graphene resistance at zero bias. Because our measurements are two-probe, quantum Hall effect-like features are strongly dependant on the aspect ratio and the geometry of the sample. As reported by Abanin et al.\cite{Abanin, Williams}, multilayer graphene samples close to a square geometry ($L\approx W$) present a conductance that is a monotonic function of the filling factor $\nu$ with marked QHE plateaus, in consistency with the data for our close to square sample ($L/W=0.7$).
Despite the fact that our devices lacked a gate which precluded the tuning of the filling factor ($\nu=nh/Be$) through the electronic density, the equivalent effect was achieved by varying the magnetic field. 

As strain was imposed on the sample new QHE features appeared. Despite the sample being 6 - 7 layers, the effect is qualitatively the same as the theoretically calculated outcome of adding domain walls to a bilayer graphene (Figure \ref{Sinisa1}). Data in Figure \ref{R_B_V-2} was collected in the following order: $80\ V$, $0\ V$, $40\ V$. The consistency of the increasing tendency of the QHE features for larger voltages on the actuators despite the data not being taken in an incremental order, testifies the reversibility of the strain imposed on the sample. We present the data side by side with the outcome of the calculations, where disorder has been included (See Supplementary Materials for details) showing a qualitative agreement. Figure \ref{DeltaR} shows the difference in the calculated magnetoresistance for a bilayer graphene with and without a domain wall. We find again an agreement with the difference of the magnetoresistance measured in the presence and absence of strain in the multilayer graphene. The qualitative accord is refined when disorder is considered in the calculations as observed in Figure \ref{DeltaR}. We observed that the difference in the magnetoresistance data with $40\ V$ on the actuators and no voltage on the actuators presented the same behavior as well as middle-way values with respect to the corresponding data for $80\ V$ on the actuators, supporting the idea that larger strain induces more domain walls on the sample. Comparison between the data and the calculations allows us to estimate the electronic density in our sample. We identified the magnetic field corresponding to the valleys in the experimental data in Figure \ref{DeltaR} with the series of filling factors associated to the calculated QHE plateaus added by the domain wall (indicated in the x-axis of the top plots in Figure \ref{DeltaR}). The advantage here is that we don't need to know the number of occupied Landau levels to extract a value for the electronic density. Even though we can infer from the theoretical calculations presented in Figure \ref{DeltaR} that disorder broadens the Landau levels (resulting in a shifting the location of the valleys in the data) we were able to estimate an electronic density of $\approx 9\times10^{11}/$cm$^2$, which is a reasonable value for our device. 

Fig. \ref{dIdV_V_B} shows the differential conductance for different voltages on the actuators at different magnetic fields, showing a discrete modulation of the current. Measurements were taken in the following order: (1) we applied different driving voltages to the MEMS actuators to tension the sample to different strain levels; (2) at each fixed driving voltage, we applied different magnetic fields; and (3) at each fixed magnetic field, we performed differential conductance measurements. (See supplemental materials for the complete set of data). Remarkably, the differential conductance measured for different voltages on the actuators was identical unless a magnetic field was imposed. The effect became observable at around $1.65\ T$, close to the onset of quantum oscillations, as observed in the magnetoresistance. As the magnetic field increased, we noticed that the curves divided into three groups: \{$0\ V$, $20\ V$\}, \{$40\ V$\} and \{$60\ V$, $80\ V$\} (see Fig. \ref{dIdV_V_B}). We again attribute these discrete changes in the conductance to new domain walls induced in the multilayer graphene, triggered by strain. 
The differential conductance with no voltage on the actuators represents a starting configuration of the domain walls in the sample. As shown in Fig. \ref{dIdV_V_B}, applying $20\ V$ doesn't impose a strong enough strain to create a new domain wall. Only  $40\ V$ and $60\ V$ generate dislocation reactions that result into discrete changes in the conductance across the sample. We interpret this behavior as strain inducing domain walls in the sample one by one. The switching effect is revealed only in the presence of a magnetic field, starting between $1.25\ T$ and $1.65\ T$, corresponding to the onset magnetic field for quantum oscillations as observed in the magnetoresistance (Figure \ref{R_B_V-2}).
The overall evolution of the features in the differential conductance near zero bias at different magnetic fields is dictated by the behavior in the absence of strain (see Figure \ref{dIdV_B}) showing that the domain walls generated by strain have little effect on the phonon modes of the suspended multilayer graphene. As detailed in the Supplementary Materials, we have found that in the absence of strain and magnetic field, the differential conductance shows features with a periodicity of $\approx 20meV$ (inset of Figure \ref{dIdV_B}) indicating a coupling to the lowest energy optical phonon mode in graphite, corresponding to neighboring non-equivalent planes vibrating in phase opposition along the c axis.

In conclusion, we have found evidence of the effect of domain walls in the QHE of suspended multilayer graphene, triggered by strain. Through a reversible and controlled adjustment of uniaxial strain, we were able to tune the appearance of domain walls, manifested as new QHE plateaus in consistency with our calculations. We observed a discrete modulation of the differential conductance activated by strain, in accordance with the tuning of the dislocation landscape in the sample. In the absence of strain, we identified features in the differential conductance corresponding to the coupling of an optical phonon mode in graphite, in agreement with previous works on suspended multilayer graphene.
Our results put in evidence that while adding regular disorder to multilayer graphene preserves its characteristic QHE features, it is only through the addition of domain walls that the quantization of the resistivity is modified. Our work constitutes a first illustration of the effect of the arrangement of domain walls in multilayer graphene in the quantum Hall effect regime, and paves the way for the use of reversible topological defects for electronic applications.

\begin{figure}
\includegraphics[width=9cm]{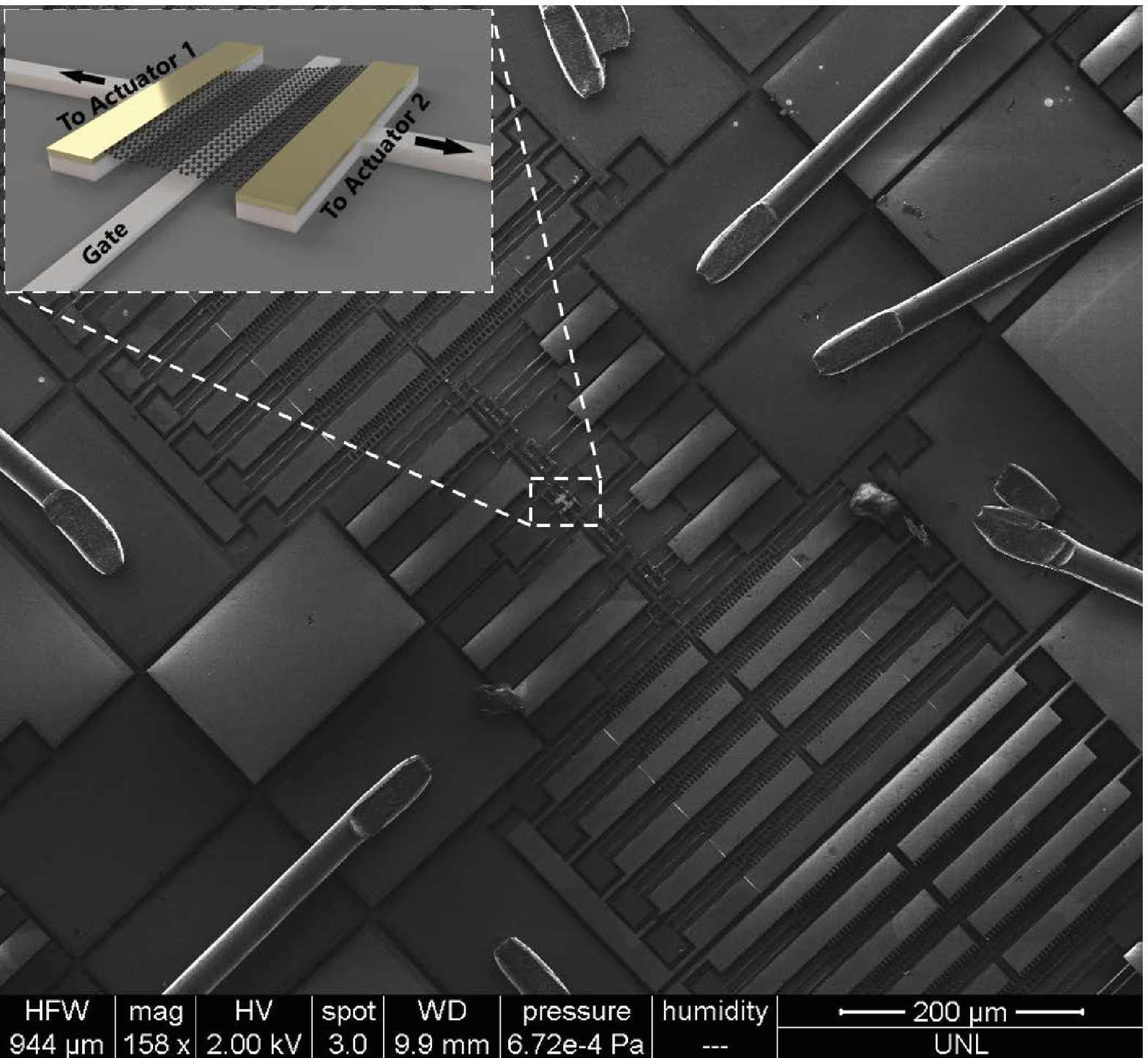}
\includegraphics[width=6cm]{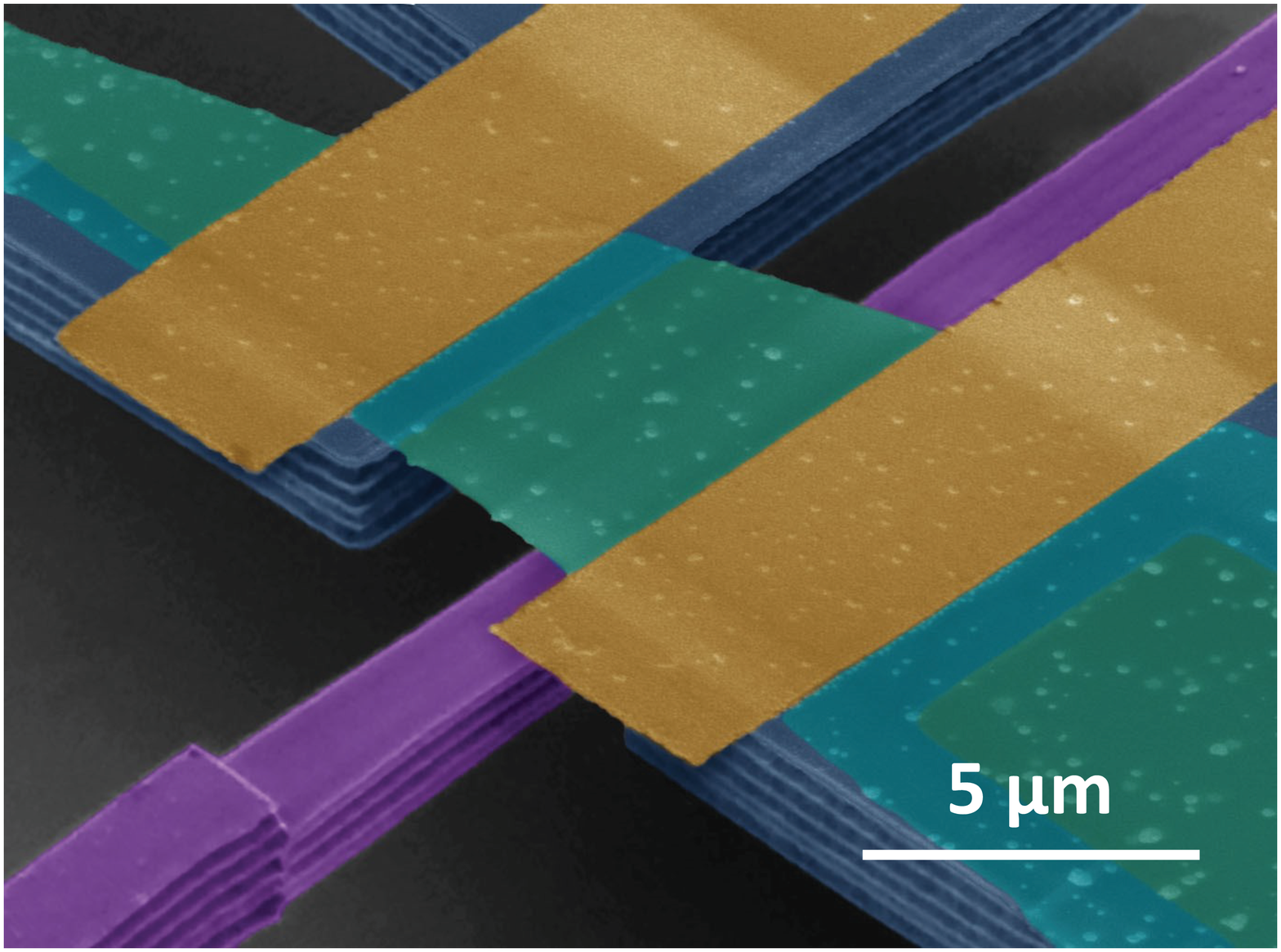}
\caption{\label{Sample}Left: SEM image of the MEMS chip showing the suspended graphene (represented in the inset) held by movable structural silicon beams connected to the two MEMS actuators. Right: Zoom-in image of the suspended graphene (green) showing source and drain electrodes (yellow) and back gate (purple).}
\end{figure}

\begin{figure}
\hspace{-10mm}
\includegraphics[width=8cm]{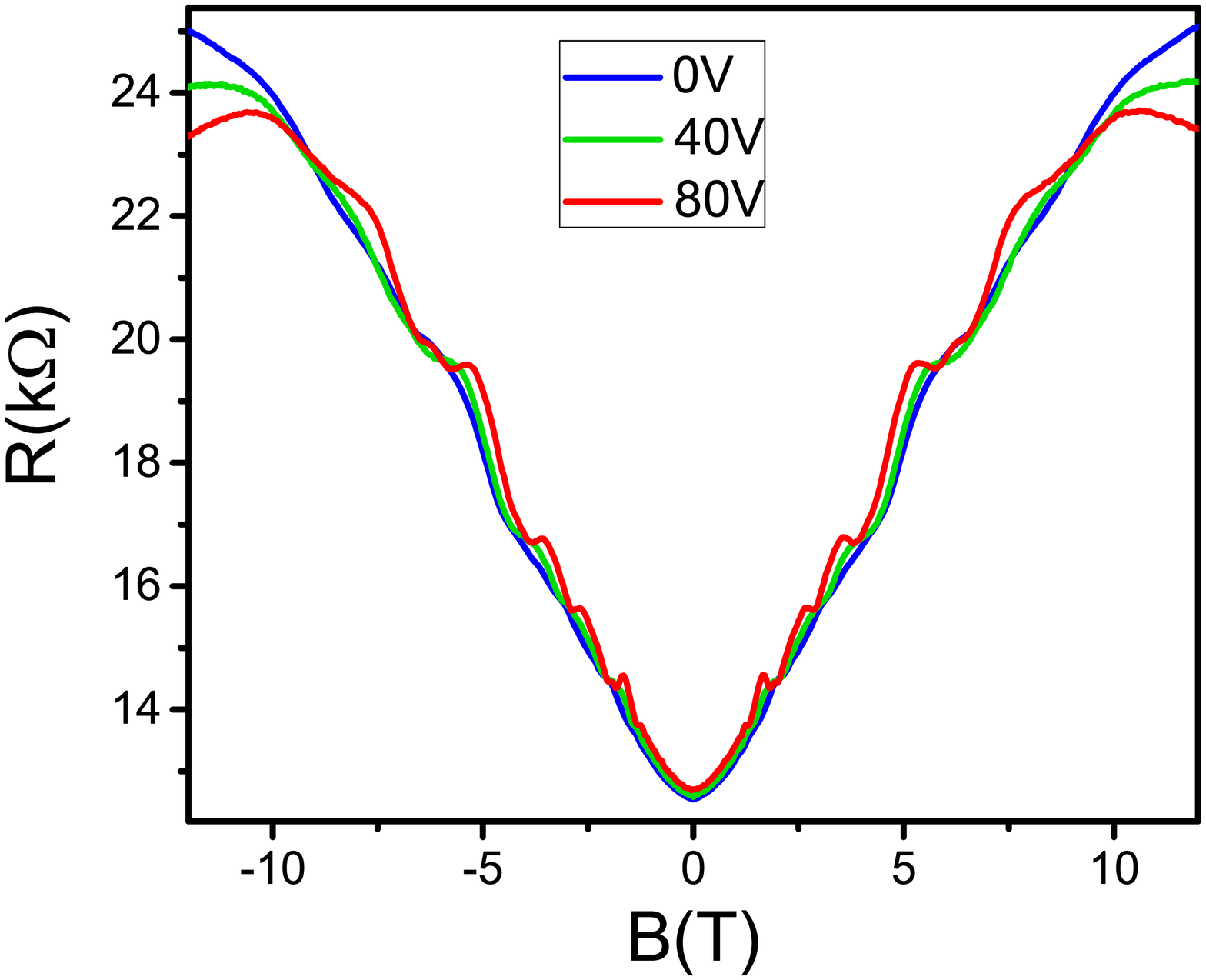}
\includegraphics[width=8cm]{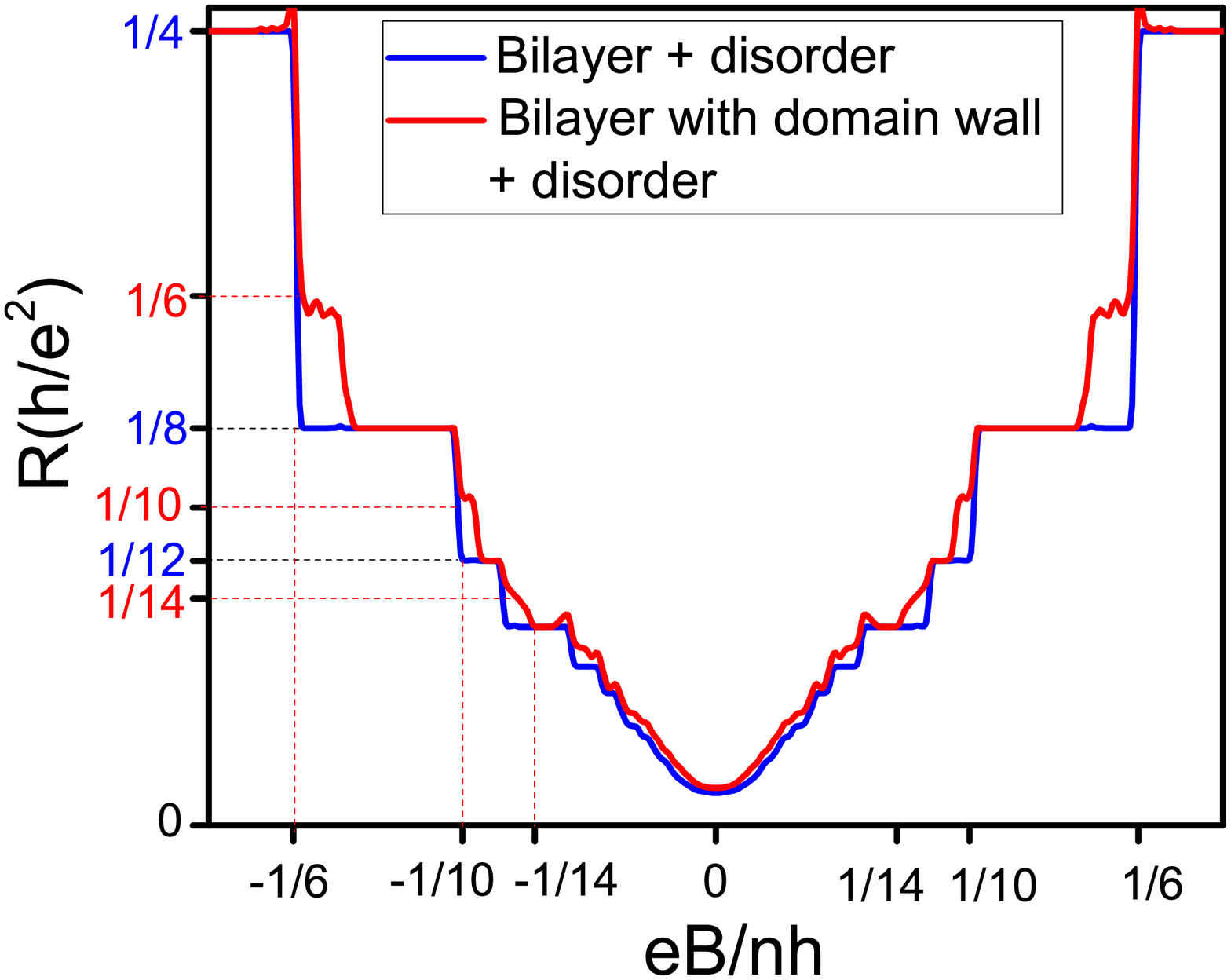}
\caption{\label{R_B_V-2} Left: Magnetoresistance at $1.5\ K$ for different voltages on the actuator. $0\ V$ represents no strain and $80\ V$ maximum strain. Right: Calculated magnetoresistance for the bilayer graphene just as in Fig. \ref{Sinisa1} now including disorder in the calculation (see Deatils in the Supplementary Materials.)} 
\end{figure}

\begin{figure}
\hspace{-10mm}
\includegraphics[width=8cm]{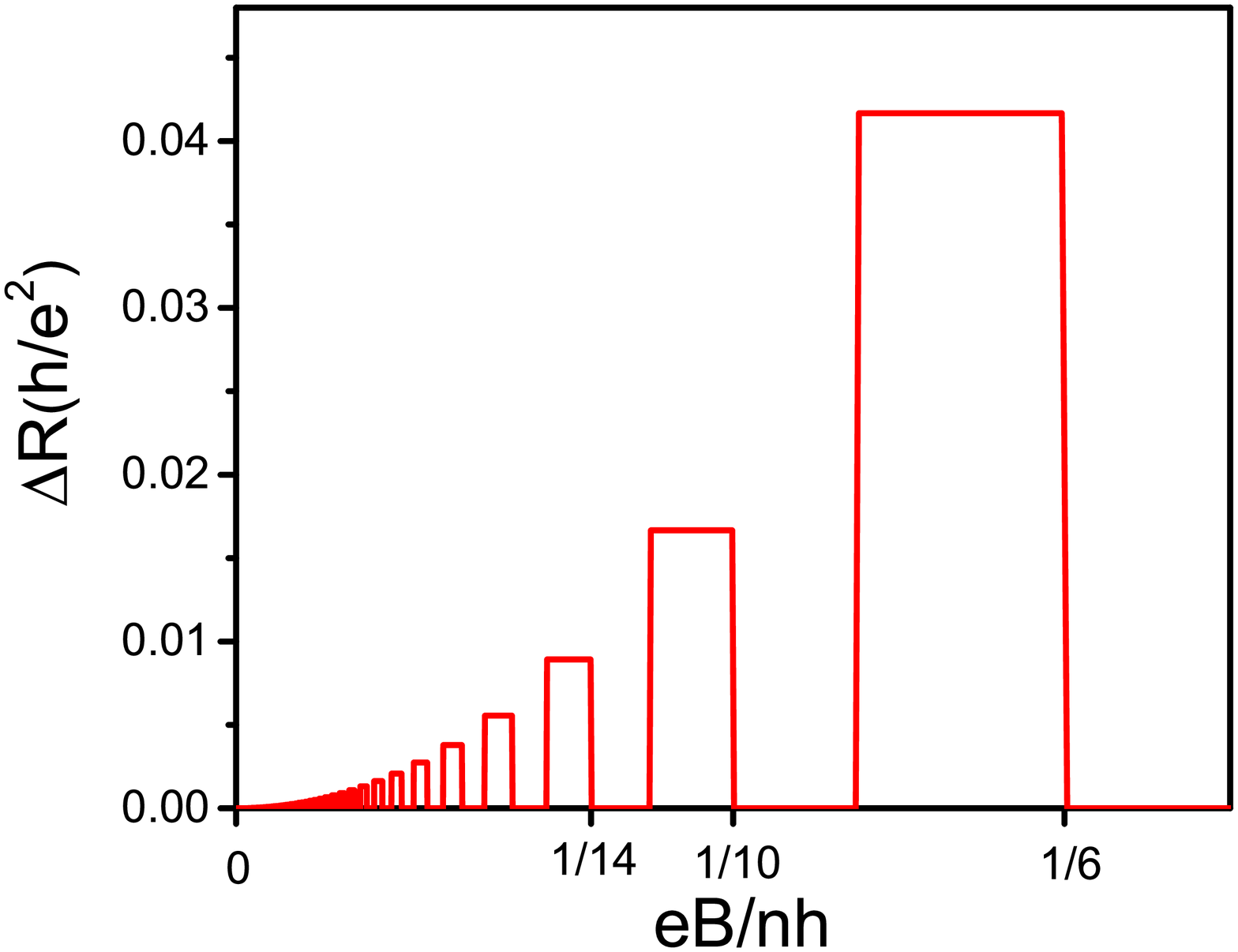}
\includegraphics[width=8cm]{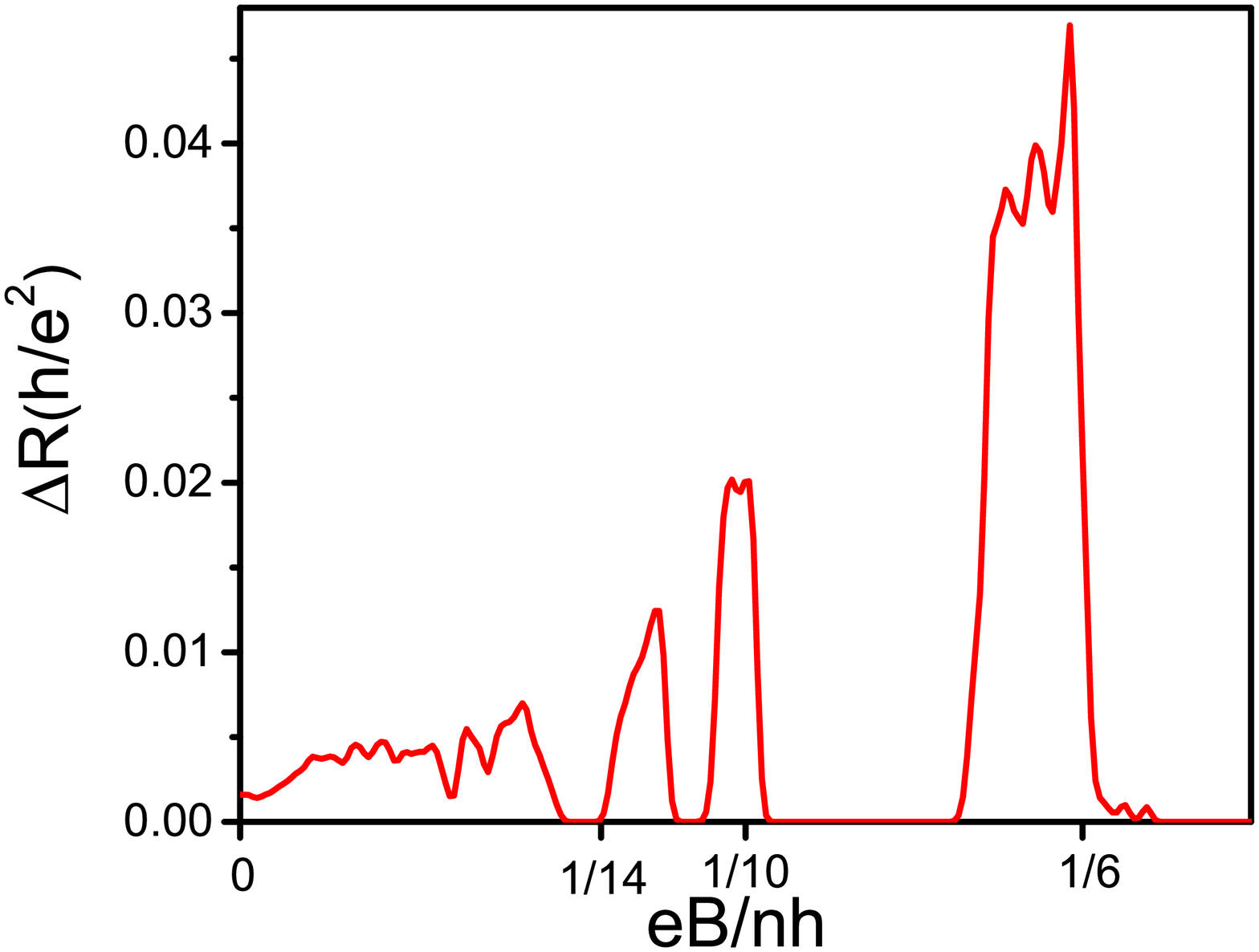}\\
\includegraphics[width=8cm]{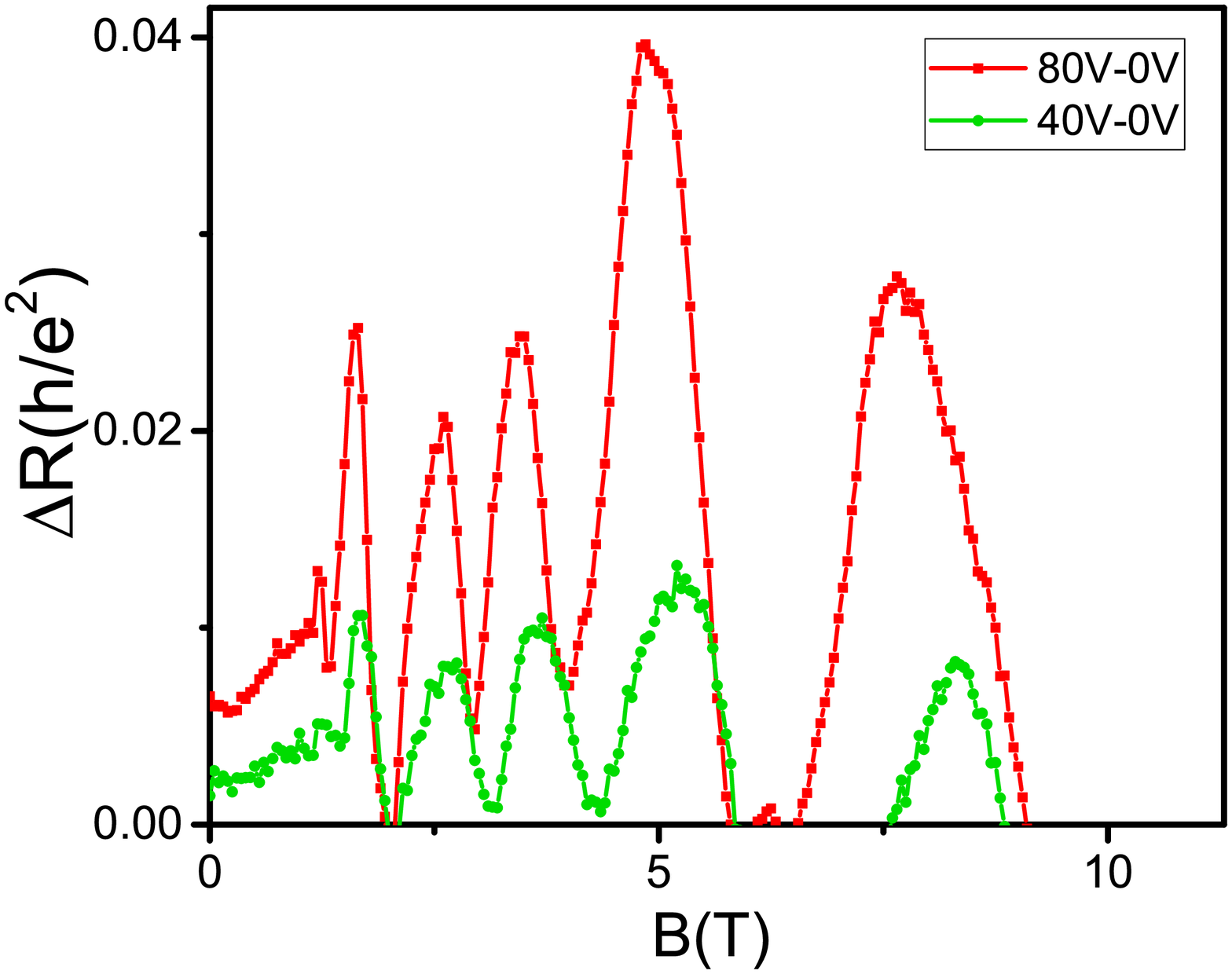}
\includegraphics[width=8cm]{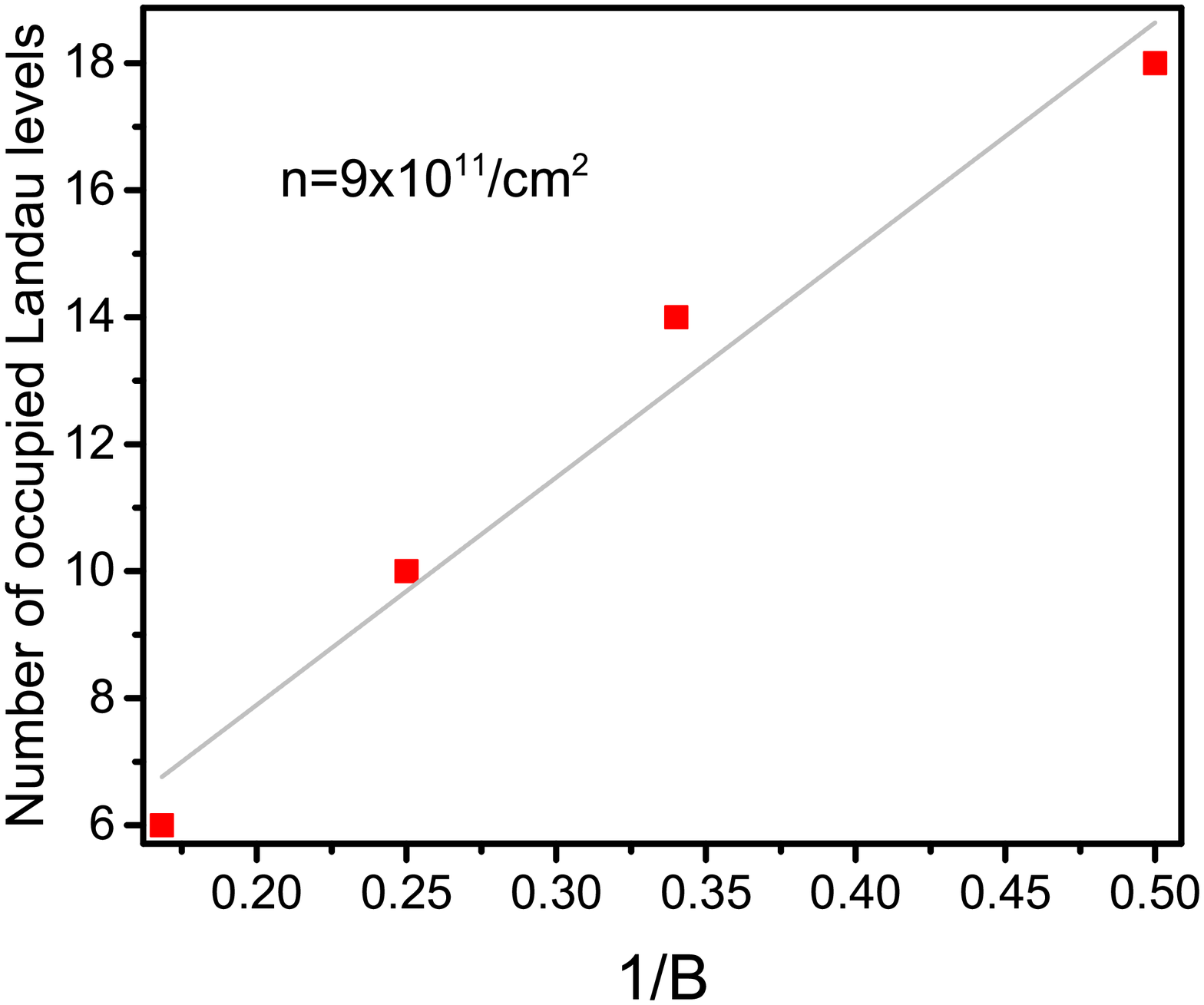}
\caption{\label{DeltaR} Difference in the magnetoresistance with and without domain walls. Top: As found from the calculations, with no disorder (left) and with disorder added (right). Bottom left: as found in the experiment. $\Delta R$ represents the difference in the magnetoresistance before and after applying $80\ V$ on the actuator (red) and $40\ V$ on the actuator (green). Bottom right: Electronic density of the device deduced from the magnetic field dependence of the filling factor identified in the experimental data (see text for details).} 
\end{figure}

\begin{figure}
\hspace{-20mm}
\includegraphics[width=80mm]{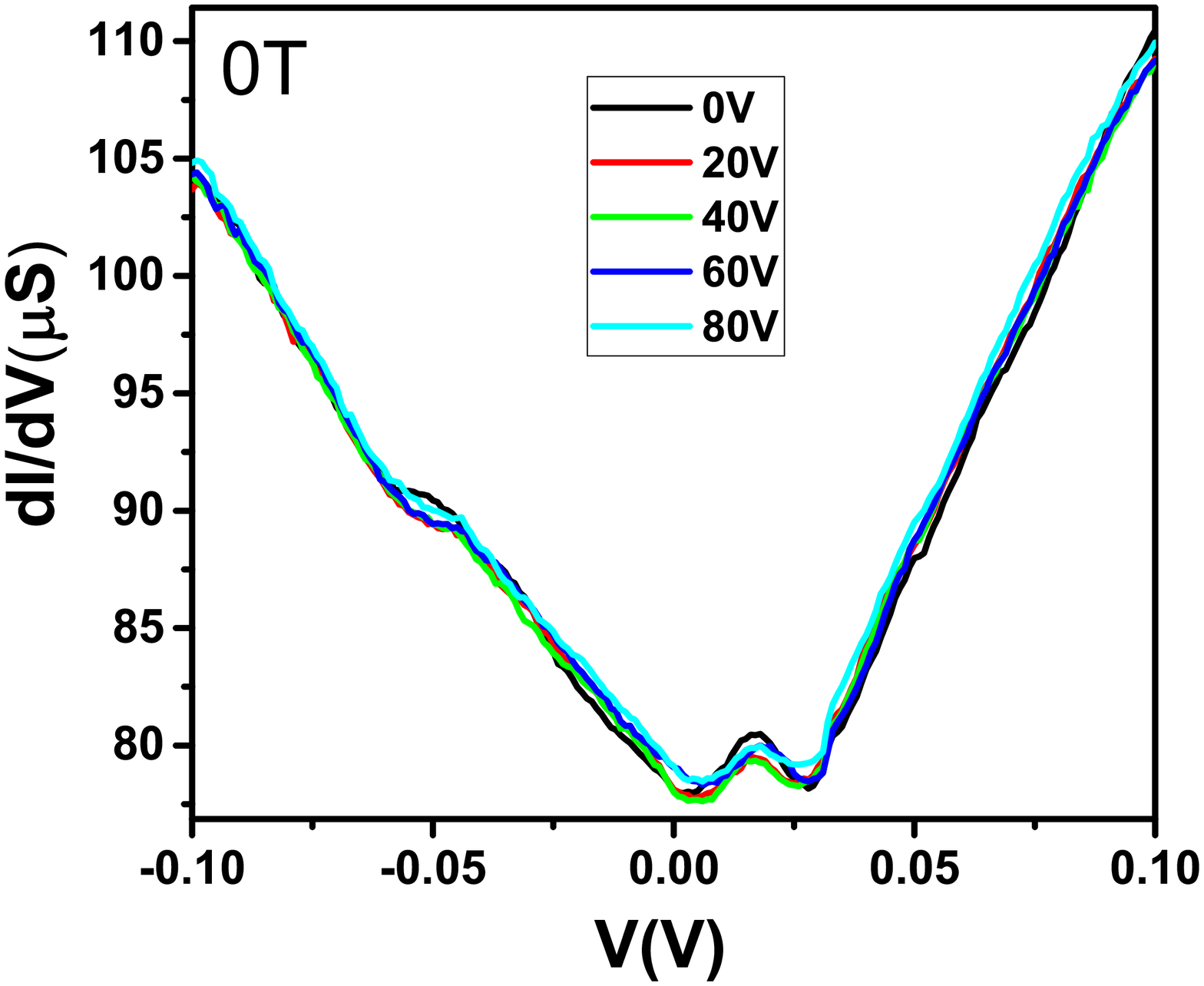} 
\hspace{-10mm}
\includegraphics[width=80mm]{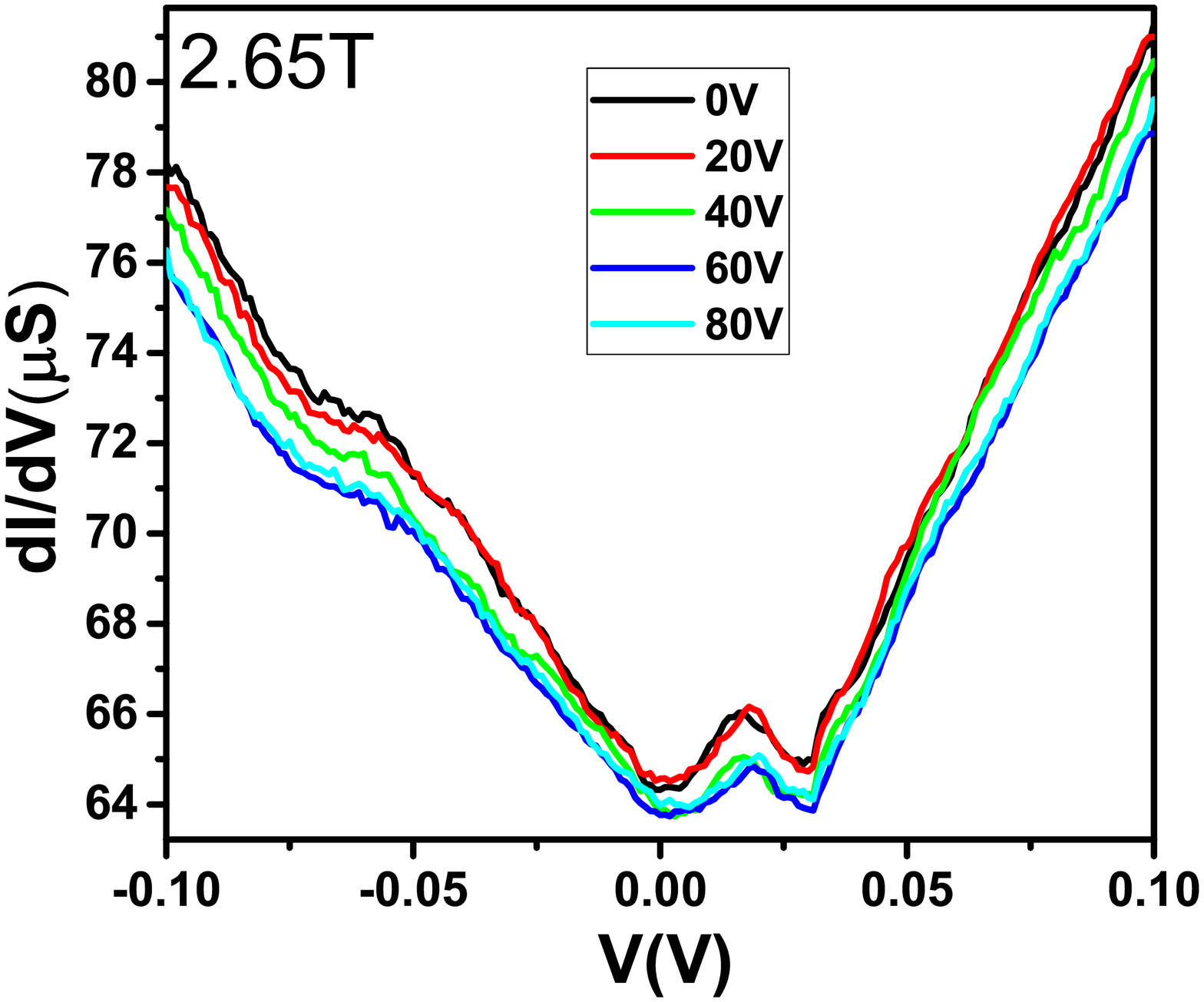}\\ 
\hspace{-20mm}
\includegraphics[width=80mm]{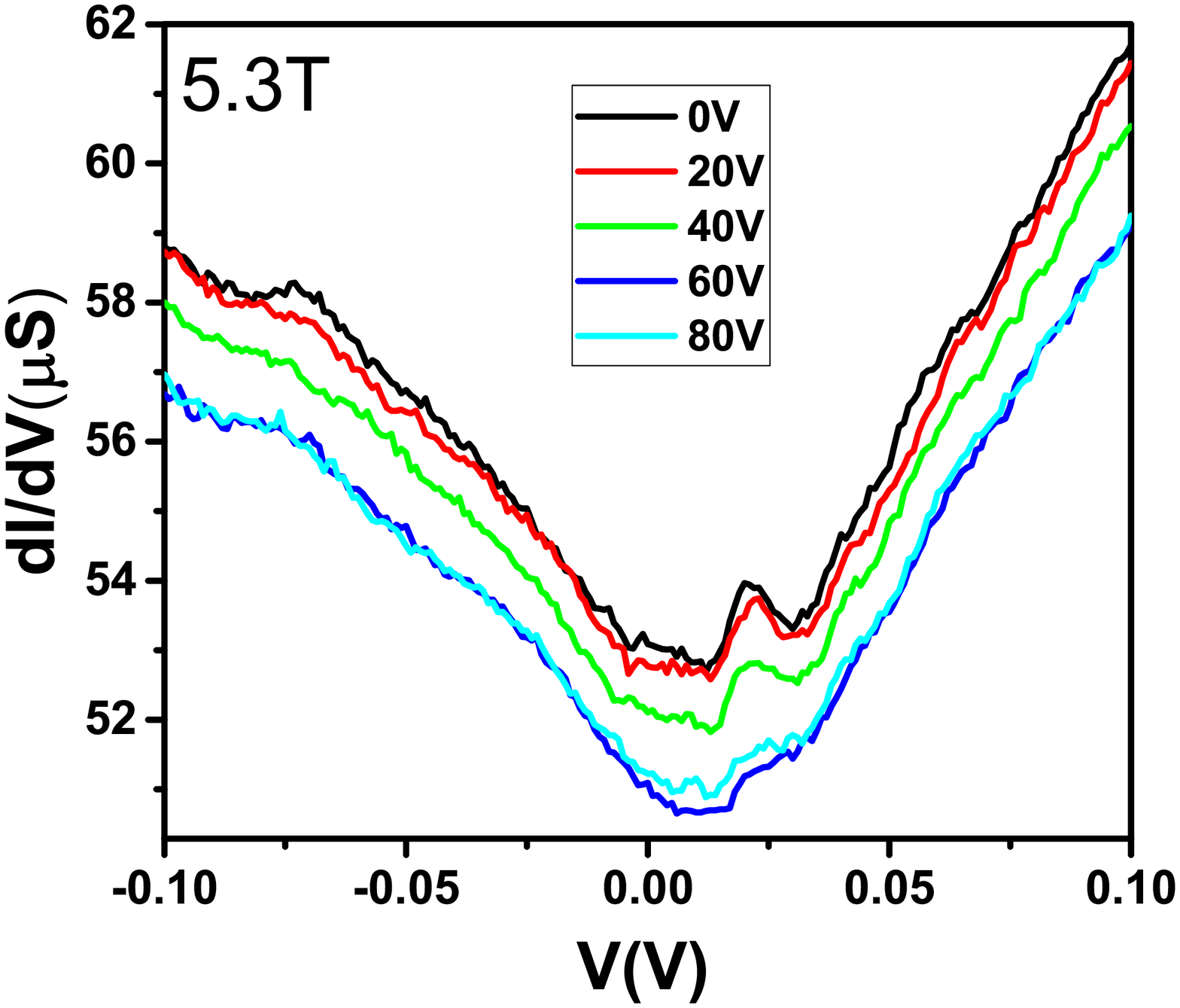} 
\hspace{-10mm}
\includegraphics[width=80mm]{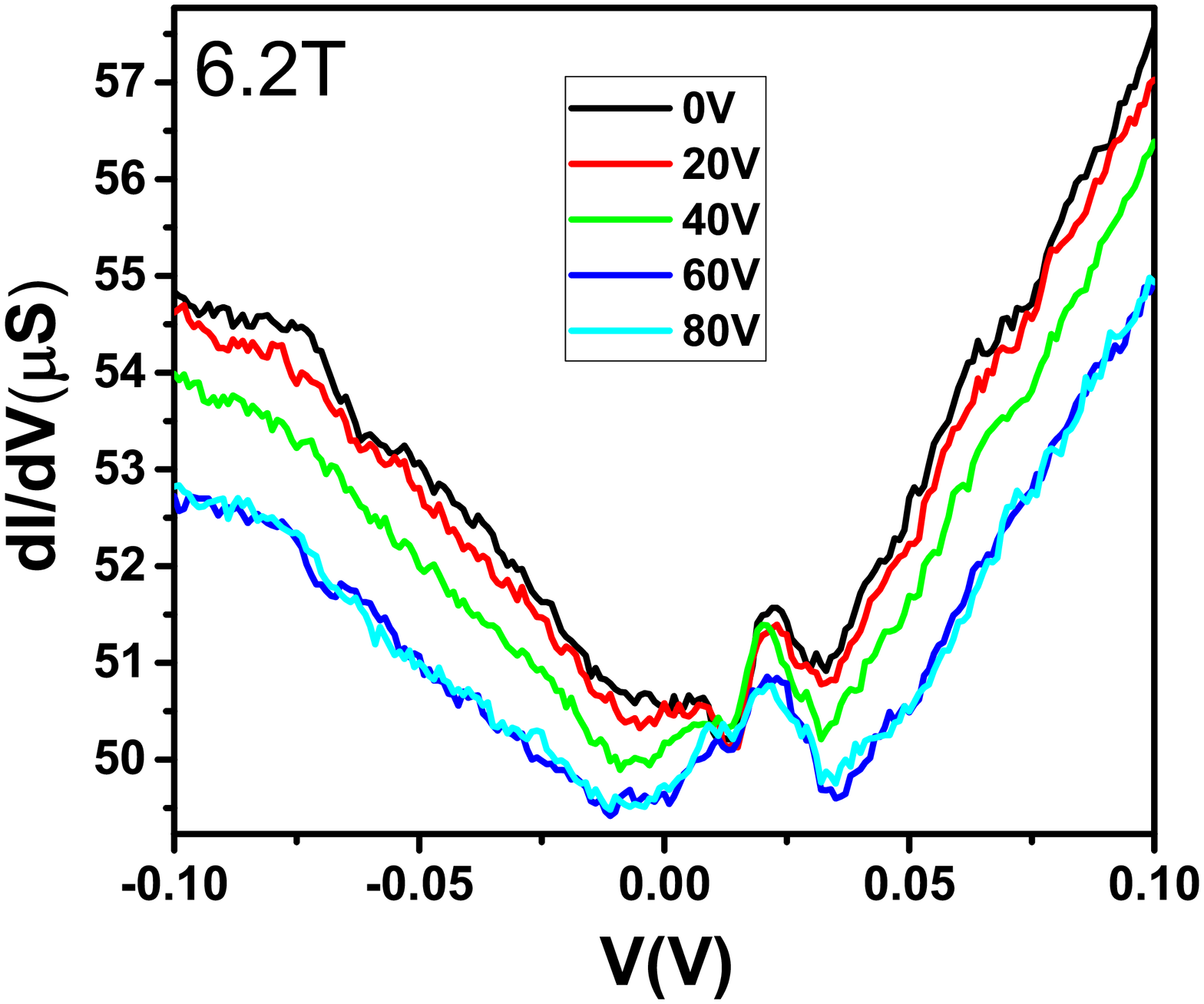}\\
\hspace{-20mm}
\includegraphics[width=80mm]{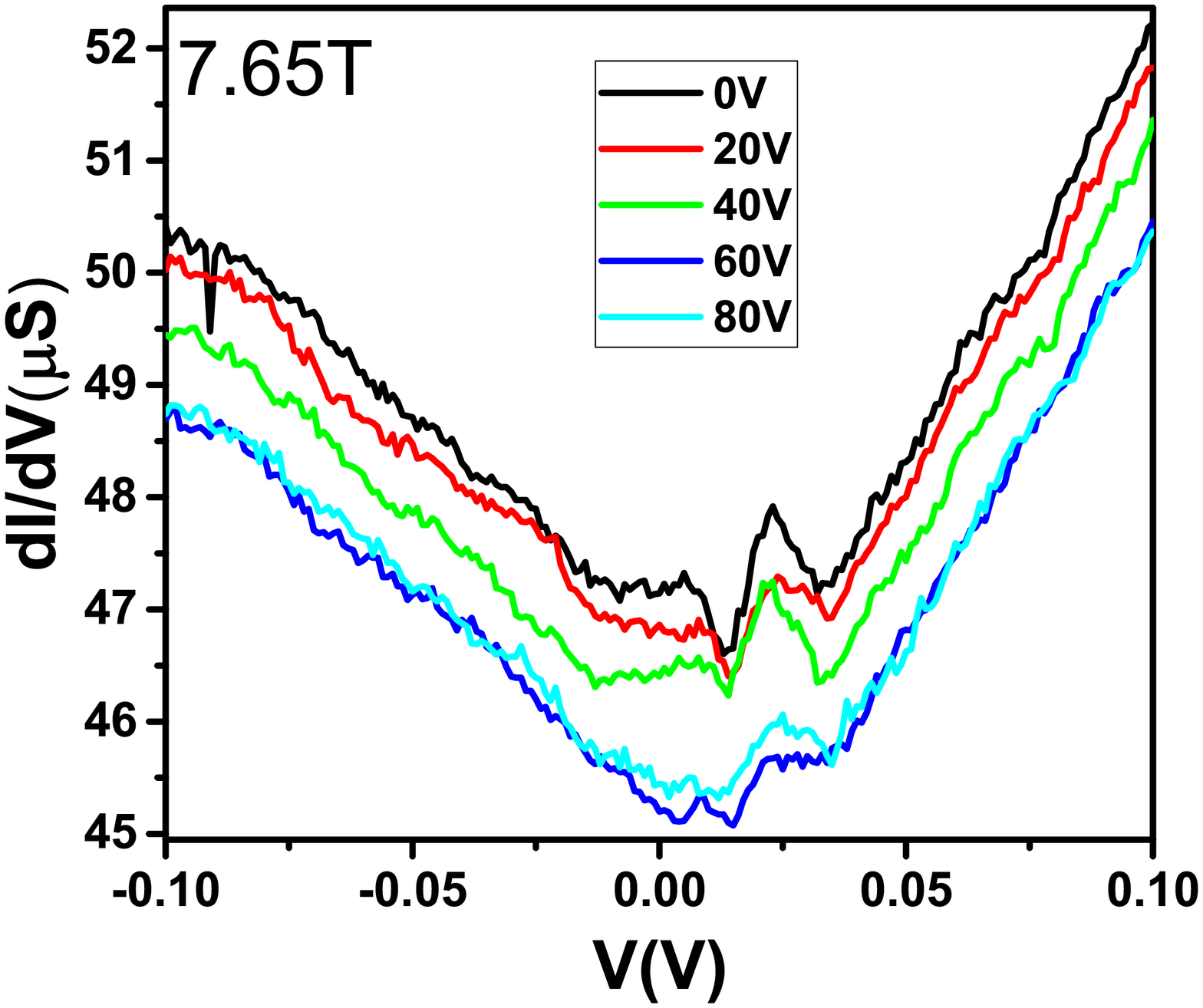} 
\hspace{-10mm}
\includegraphics[width=80mm]{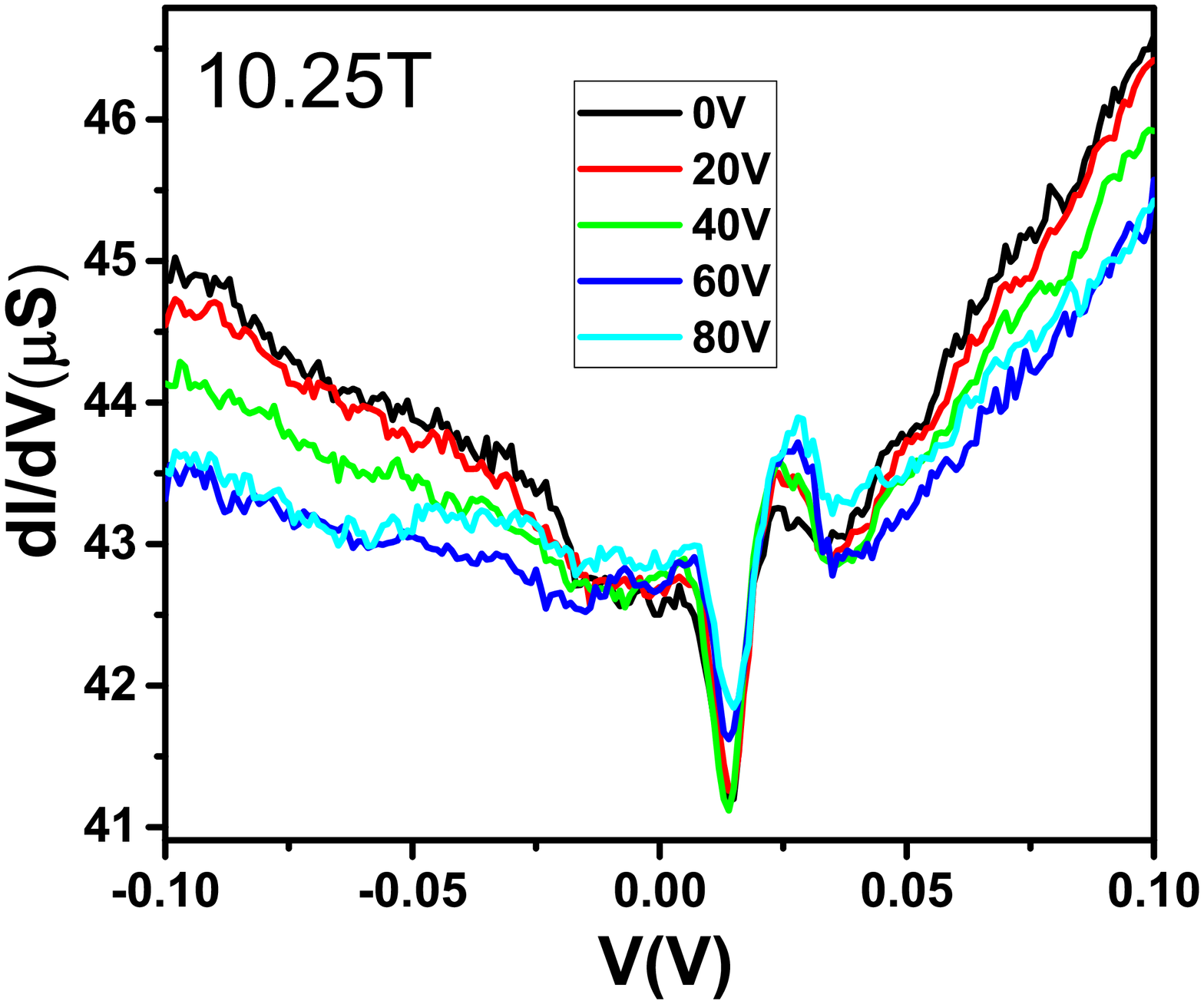}
\caption{Differential conductance as a function of the bias voltage for different voltages on the actuator. Each panel shows data taken at different magnetic fields (0T, 2.65T, 5.3T, 6.2T, 7.65T, 10.25T)}
    \label{dIdV_V_B}
\end{figure}

\begin{figure}
\includegraphics[width=10cm]{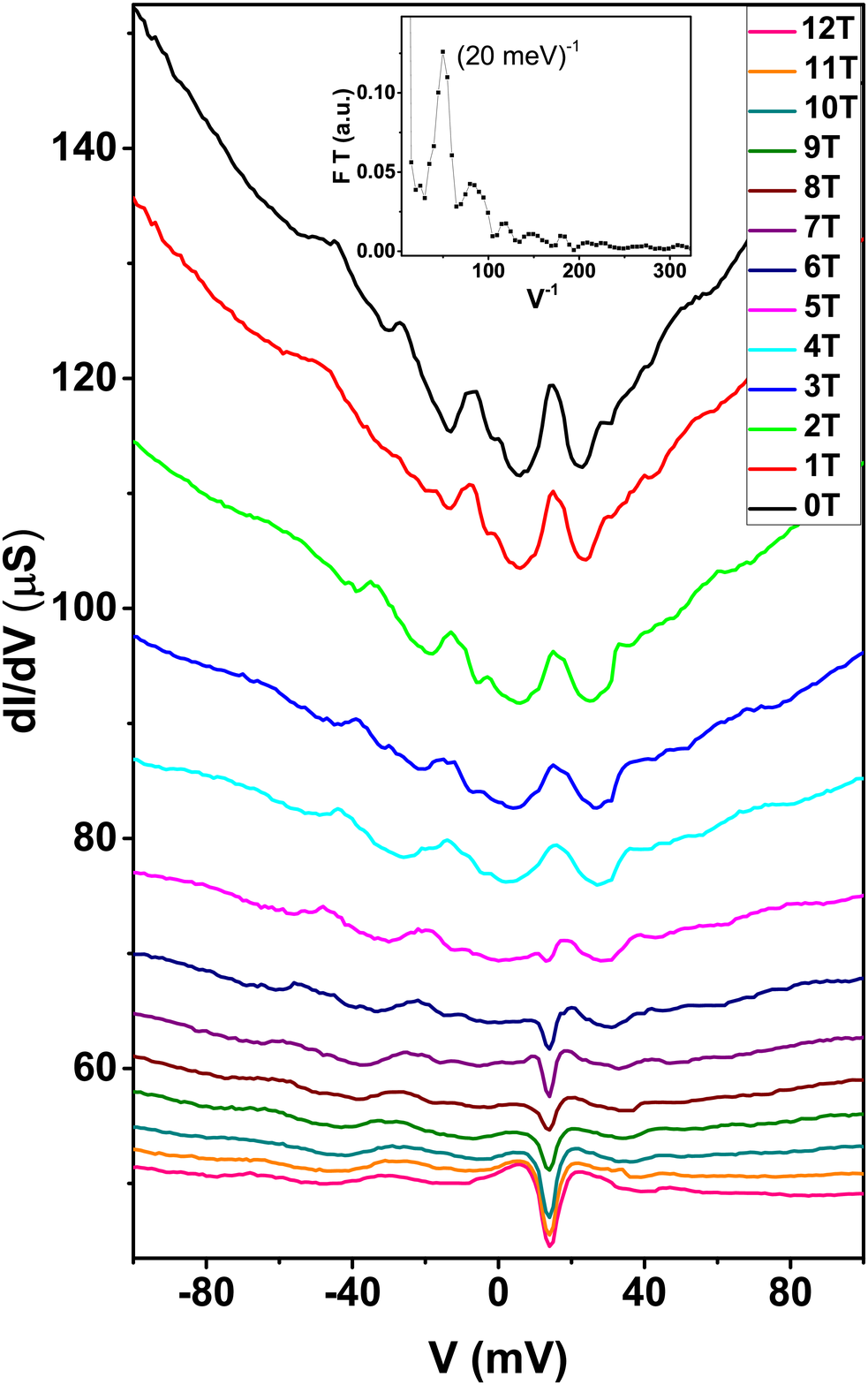}
\caption{\label{dIdV_B} Bias dependent differential conductance (dI/dV) in the absence of strain at different magnetic fields (0T to 12T). Inset: Fourier transform of the dI/dV at 0T that indicates a periodicity of $(20\ meV)^{-1}$ }
\end{figure}

\begin{acknowledgments}
Funding for this work was provided by the U.S. Department of Energy, Office of Science, Office of Basic Energy Sciences under contract DE-SC0018154. S. C. was supported by the National Science Foundation, award number DMR-1848074. CO-A would like to acknowledge invaluable advice from David Warren from Oxford Instruments. We would like to acknowledge discussions an preliminary calculations by Gautam Rai and Stephan Haas.
\end{acknowledgments}

\newpage
\section*{Supplementary Materials}

\section{Fabrication process of the MEMS $-$ graphene hybrid device}
\begin{figure} 
\includegraphics[width=16cm]{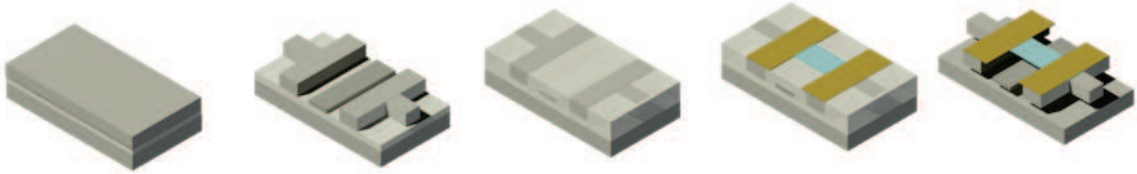}
\caption{\label{Fig1}Device fabrication flow}
\end{figure}

The process starts from SOI wafers (See Figure \ref{Fig1}). The top silicon is first patterned to the designed structures using photolithography and deep reactive ion etching. Sacrificial silicon oxide is then filled in between the spacing of the structural silicon. Chemical-mechanical polishing is used to make a smooth top surface for graphene transferring.  Thin graphene flakes are first mechanically exfoliated on a Si/SiO2 substrate. The flakes with desired number of layers are identified by its optical contrast then confirmed by its distinct Raman signatures. Then the target graphene flake is transferred with a polymer support by a micromanipulator to a designated location of MEMS chips with high accuracy.  Subsequently standard electron-beam lithography techniques are used to pattern graphene and define clamping metal electrodes.  In the final step, the silicon dioxide sacrificial filler is etched away in 5:1 BHF, and the chips are dried with a critical point drier to acquire freestanding MEMS devices. The multilayer graphene is suspended together with the structural silicon beams because the silicon dioxide beneath is fully etched away. The movable structural silicon beams are connected to two microactuators that can pull the graphene sample from both ends to opposite directions, controlling the displacement of the graphene. 

\begin{figure}
\includegraphics[width=10cm]{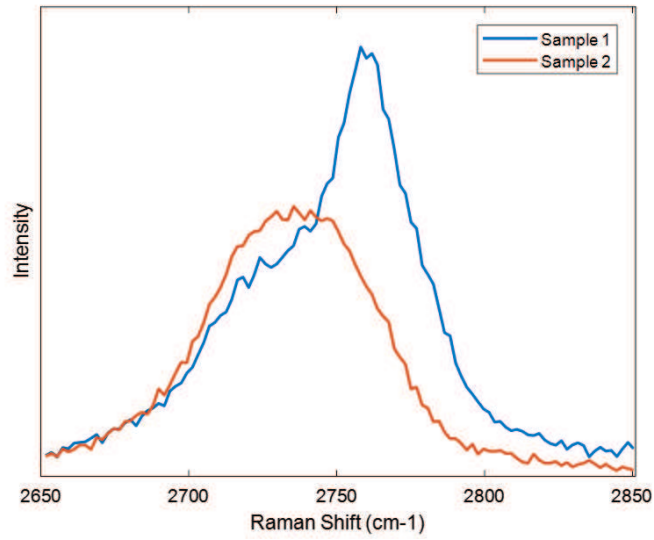}
\caption{\label{Fig2}Characterized layer number of graphene flakes by Raman spectroscopy. Sample 2 (7-layers) corresponds to the sample presented in the main text and sample 1, a three layer graphene as reference.}
\end{figure}

A library of graphene flakes with different layer numbers on 300nm SiO2 substrate identified by optical contrast is established using the approach described in literature \cite{Li}. Raman spectra of these samples are taken using a 488nm laser to build a standard library, and used as another approach to identify the layer number. The 2D peak is used for characterization. A reference sample is characterized as a tri-layer and our sample is estimated as 6-7 layer (see Figure \ref{Fig2}).

\section{Mechanical Modeling}
\begin{figure} 
\includegraphics[width=10cm]{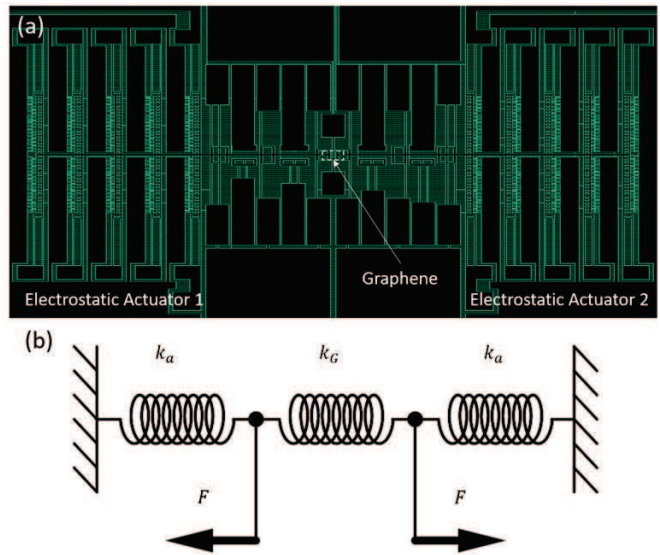}
\caption{\label{Fig3}Schematic of MEMS-graphene hybrid device. (a) Layout of the system. (b) Lumped mechanical model.}
\end{figure}

As shown in Figure \ref{Fig3}a, the MEMS-graphene hybrid device consists of 2 comb-shape electrostatic actuators at both ends  pulling the graphene flake in the middle to opposite directions. A lumped model is used to estimate the displacement of the graphene. When the two identical actuators are working at the same actuation voltage the system can be simplified as shown in Figure \ref{Fig3}b. 

\begin{figure} 
\includegraphics[width=10cm]{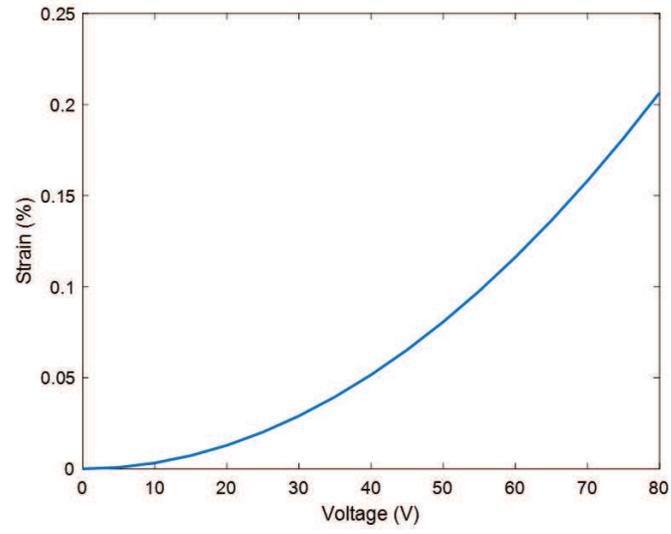}
\caption{\label{Fig4}Actuation voltage - strain relation of the graphene sample 1.}
\end{figure}

The compatibility of deformation and equilibrium of the system lead to the following equations:
\begin{eqnarray}
\nonumber F=k_aX_a+k_GX_G \\
\nonumber2X_a=X_G,
\end{eqnarray}

where $F$ is the force generated by the electrostatic actuators, $k_a$ and $k_G$ are the spring constant of the actuator and graphene flake respectively, $X_a$ and $X_G$ are the displacement of the actuator and graphene respectively. The spring constant of graphene and actuator are given by:
\begin{eqnarray}
\nonumber k_G=\frac{E_GA_G}{L_G}\\
\nonumber k_a=\frac{2E_ab^3h}{L_a^3},
\end{eqnarray}
where $E$ is young’s modulus, $A$ is the cross section area, $b$ is width, $h$ is thickness and $L$ is length. 
The force generated by the electrostatic actuators is calculated as
\begin{equation}
F=\frac{n\epsilon_0b}{d}V^2,   
\end{equation}
where $n$ is the number of comb figures in the electrostatic actuator, $\epsilon_0$ is dielectric constant, $d$ is the gap spacing between the fingers, $V$ is the actuation voltage. By solving the equations above, the actuation voltage – strain relation is found, as shown in Figure \ref{Fig4}. At $20\ V$, $40\ V$, $60\ V$ and $80\ V$, the strain of the graphene flake is estimated as $0.01\%$, $0.05\%$, $0.12\%$ and $0.21\%$ respectively.

\begin{figure} 
\includegraphics[width=10cm]{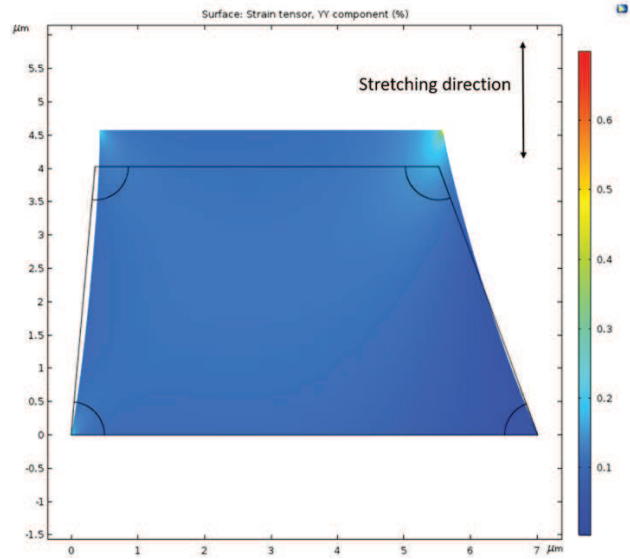}
\caption{\label{Fig5}Finite element analysis of strain distribution of the graphene sample 1 under uniaxial tension. }
\end{figure}

Because the shape of graphene is an acute trapezoid (see Figure 1 in main text), we realize the strain distribution will not be uniform. We also performed finite element analysis to simulate the strain distribution of graphene using COMSOL. As shown in Figure \ref{Fig5}, when the actuation voltage is at 40V, even though the average strain is small, we can clearly see strain concentration at the corner, which we believe are the places where dislocations are generated first. 

\section{Theoretical calculations}
We computed the quantum transport through the bilayer graphene with a domain wall using the open-source computer package Kwant \cite{Kwant}. For that purpose we considered the bilayer graphene of the finite width of 12 nm.  This width is large enough to eliminate finite-size effects.  The scattering region of the bilayer graphene has a length of 7.3 nm.  On each side of the bilayer graphene, the scattering region is connected to semi-infinite leads.  The tight-binding model for graphene is based on Slater-Koster parameterization from Ref. \cite{Trambly}. Since Slater-Koster parameterization depends only on distances, and angles, between atoms, we can apply the same parameterization both to the region away from the domain wall, or the region near the domain wall.  We applied the magnetic field to the model by using the Peierls substitution.  The domain wall was modeled by shifting one of the layers of graphene by a fraction of a unit-cell vector so that on each side of the scattering region there is a continuous transition into the semi-infinite leads.  The domain wall is modeled with a sigmoid function $\frac{1}{e^{x/a}+1}$ where $x$ is location along the bilayer and parameter $a$ equals 0.8 nm. However, we tested that a larger domain wall-size gives qualitatively the same final result for transport. The disorder was modeled by adding a constant onsite energy shift to every carbon site.  The energy shift was chosen at random from a Gaussian distribution with a full-width-at-half-maximum of 0.3 eV. Transport was computed averaged over ten realizations of the disorder.

\section{Differential conductance with no strain}
\begin{figure}
\includegraphics[width=8cm]{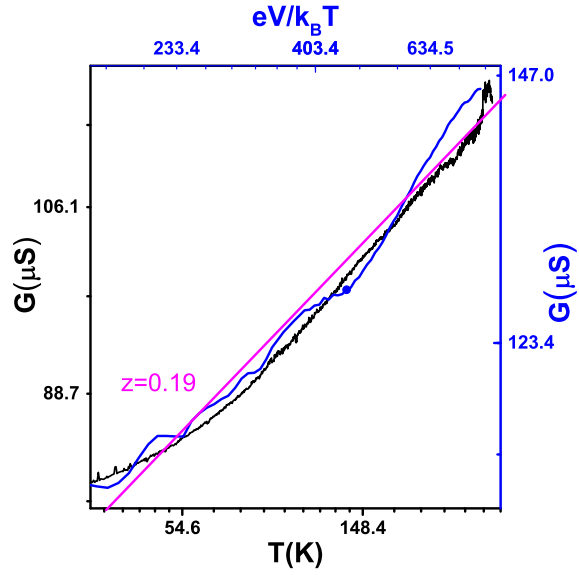}
\caption{\label{GvsT}Temperature dependence of the conductance (left-Bottom axis) compared to the bias-dependent differential conductance (right-top axis) in a double logarithmic scale. Purple line shows a fit to equation \ref{powerLaw} with exponent $z=0.19$}
\end{figure}

In the following, we present  results in the absence of strain or magnetic field. The bias dependent differential conductance revealed a V shape, consequence of the linear density of states in graphene with energy \cite{Castro-NetoRev}. Typically, access to the density of states through electronic transport is granted by tunnel contacts. In our samples, the temperature dependence of the resistance as well as the bias dependent differential conductance exhibited a power law, characteristic of Coulomb blockade through a tunnel junction \cite{Ingold-Nazarov, Alexei}:
\begin{equation}\label{powerLaw}
dI/dV=T^zf(eV/k_BT)
\end{equation}
with $\lim_{x\rightarrow 0}f(x)=C$ where C is a constant and $\lim_{x\rightarrow\infty}f(x)=x^z$. Figure \ref{GvsT} shows the temperature dependence of the conductance as well as the bias dependence for the sample shown in  the main text at 1.5 K, represented in a double logarithmic scale. The exponent found ($z=0.19$) is similar to values reported for multiwall carbon nanotubes \cite{Bachtold, Bockrath_NT} and suspended multiple layer graphene with low conductance contacts \cite{Alexei}. 

The differential conductance presented additionally an anomaly at zero bias. Zero bias anomalies have been associated in tunneling experiments to an interplay of electronic correlations and disorder \cite{Altshuler_Aronov, Imry_PRL}. Here, we observe replica of the zero-bias feature at multiples of $\approx20\ meV$, as seen in Figure 6 of the main text for $B=0$ (black curve). Such behavior was observed in suspended multilayer graphene by Chepelianskii et al. \cite{Alexei} and was associated to dynamical Coulomb blockade, where tunneling takes place through a small capacitance formed at the electrodes in series with an ohmic electromagnetic environment (the graphene foils) that exchanges energy with the tunneling quasiparticles, on a scale that is significantly smaller than the charging energy of the tunneling capacitors \cite{Ingold-Nazarov}. The periodicity of the anomalies in our data is confirmed by Fourier transform and shown as an inset in Figure Figure 6 of the main text. Following ref. \onlinecite{Alexei}, we associate these features to replica of the Coulomb blockade anomaly, consequence of the coupling of the dissipating electromagnetic environment to the lowest energy optical phonon mode in graphite (ZO') \cite{Vitali, Wirtz} that results into an oscillating transmission of the barriers at the contacts \cite{Alexei}. The ZO' optical phonon mode in graphite emerges from the out-of-plane acoustic phonon mode in graphene (ZA) \cite{Wirtz} and corresponds to neighboring non-equivalent planes vibrating in phase opposition along the c axis \cite{Alexei}. 
The similarity between the graphite phonon mode observed here and the one reported in ref. \onlinecite{Alexei} is likely related to the comparable aspect ratios of the samples ($L<W$). 
The impact of the samples' aspect ratio is also revealed in the magnetoresistance data, as detailed in the main text.

Features in the differential conductance were highly influenced by the presence of a vertical magnetic field, as observed in Figure 6 of the main text. Replica of the zero bias anomaly evolved into a single dip at high enough fields ($\approx5\ T$). A similar effect was observed in ref. \onlinecite{Alexei} and attributed to the field-dependent density of states in the multilayer graphene. The periodicity of the features ($20\ meV$) has a characteristic frequency $\omega_o$ comparable to graphene's cyclotron frequency at $1\ T$, assuming that the Fermi energy is near the Dirac point. 

Thermal cycling created a non-negligible effect on the sample resistance as well as on the differential conductance features. Following the measurements shown in Figure 6 of the main text, the sample went through a thermal cycle (to room temperature and back to $1.5\ K$, as shown in Fig. \ref{GvsT}) and a cycling of the voltage imposed on the actuators ($0\ V$ to $80\ V$). Impact of these experiments is found by comparing the data in Figure 6 of the main text for $B=0\ T$ and the first panel in Fig. \ref{dIdV_V_B}, where the differential conductance taken under similar conditions is represented. While there are similar features, some of the zero bias replica have been lost. It has been reported that thermal cycling can create, move and annihilate stacking boundaries across bilayer graphene \cite{San-Jose}. Here, we believe that a similar effect is responsible for the changing differential conductance. 


\section{Effects of strain on differential conductance measurements}
\begin{figure}
\hspace{-10mm}
\includegraphics[width=60mm]{dIdV_0T.eps} 
\hspace{-10mm}
\includegraphics[width=60mm]{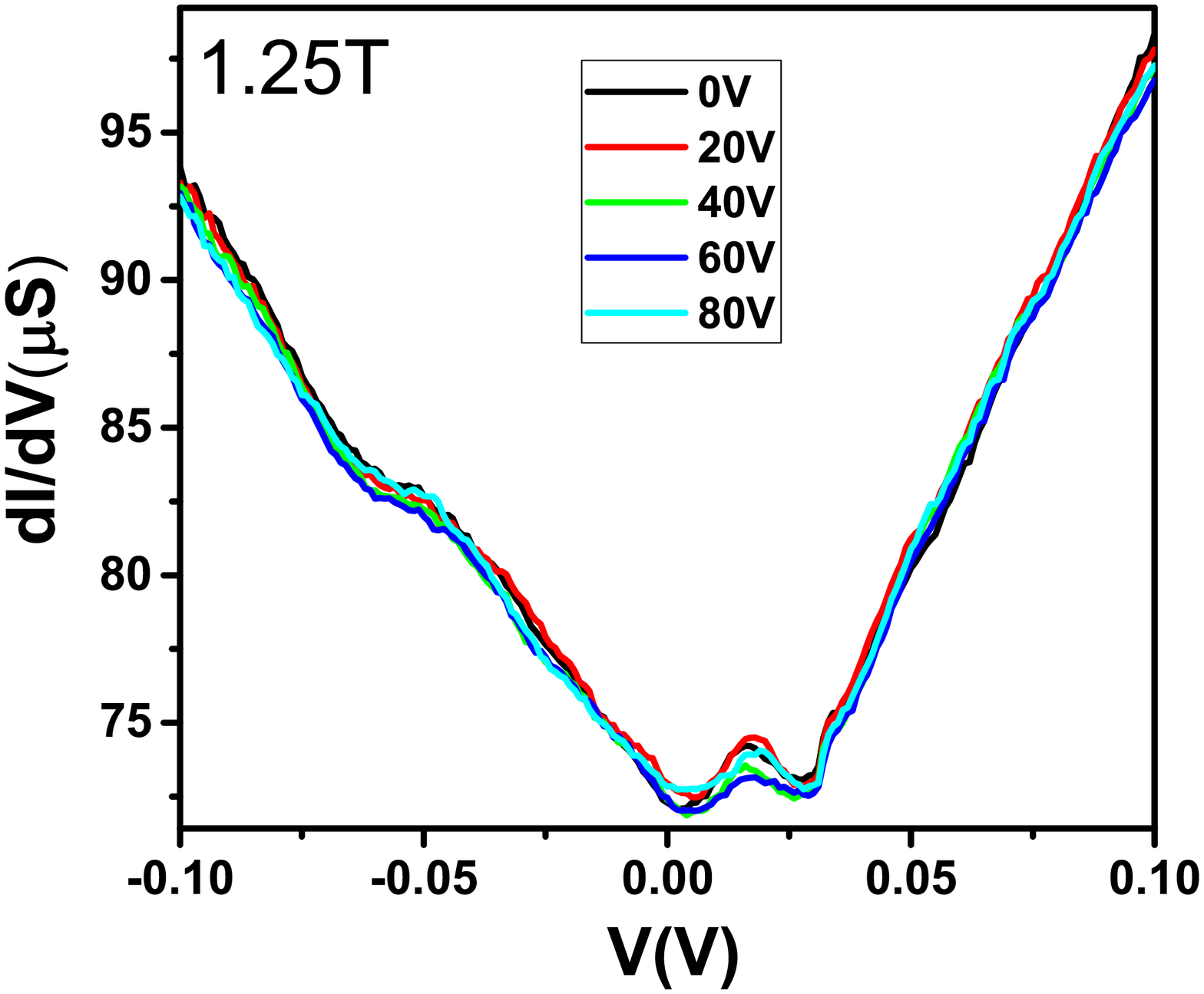}
\hspace{-10mm}
\includegraphics[width=60mm]{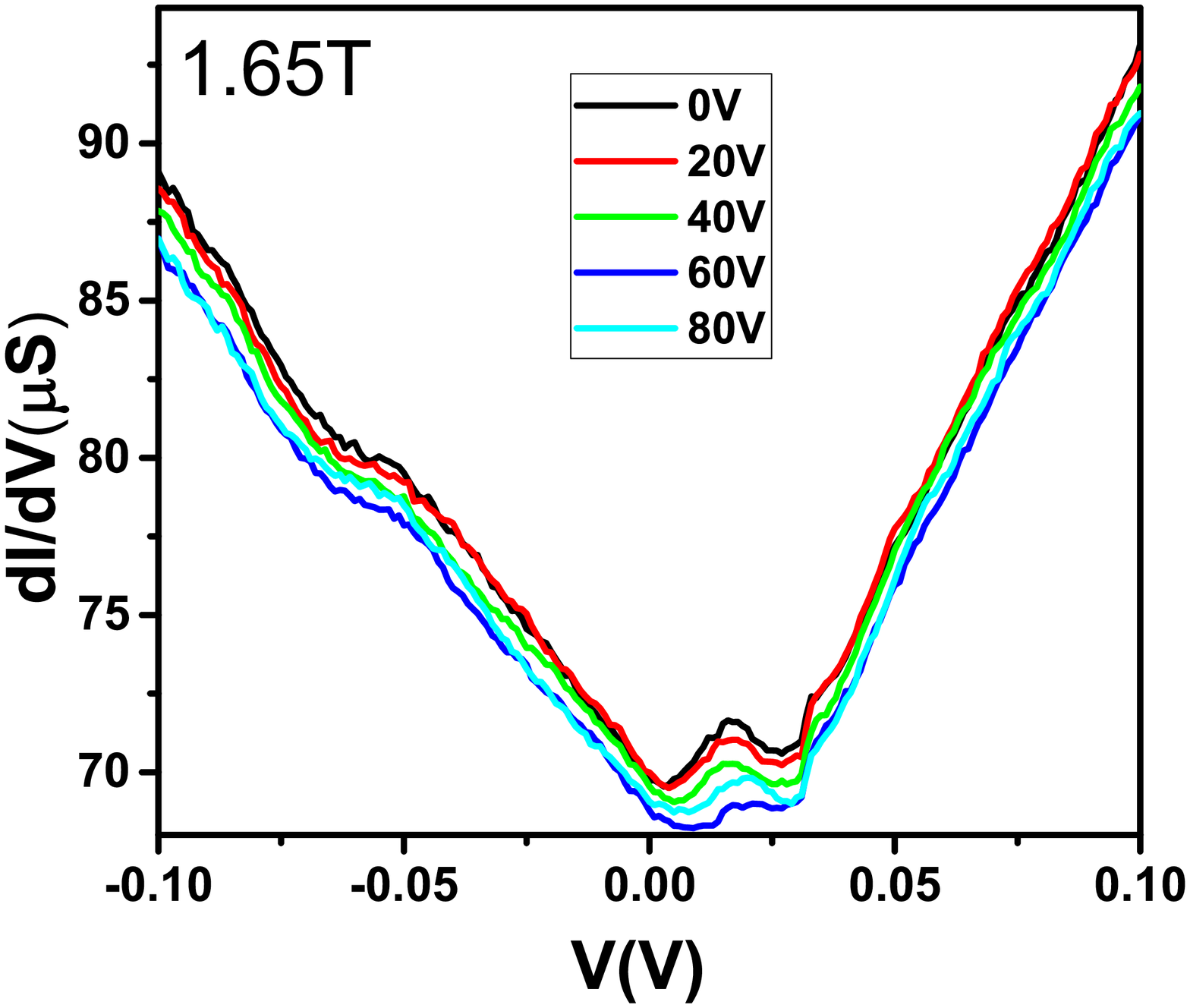}\\ 
\hspace{-10mm}
\includegraphics[width=60mm]{dIdV_2p65T.eps}
\hspace{-10mm}
\includegraphics[width=60mm]{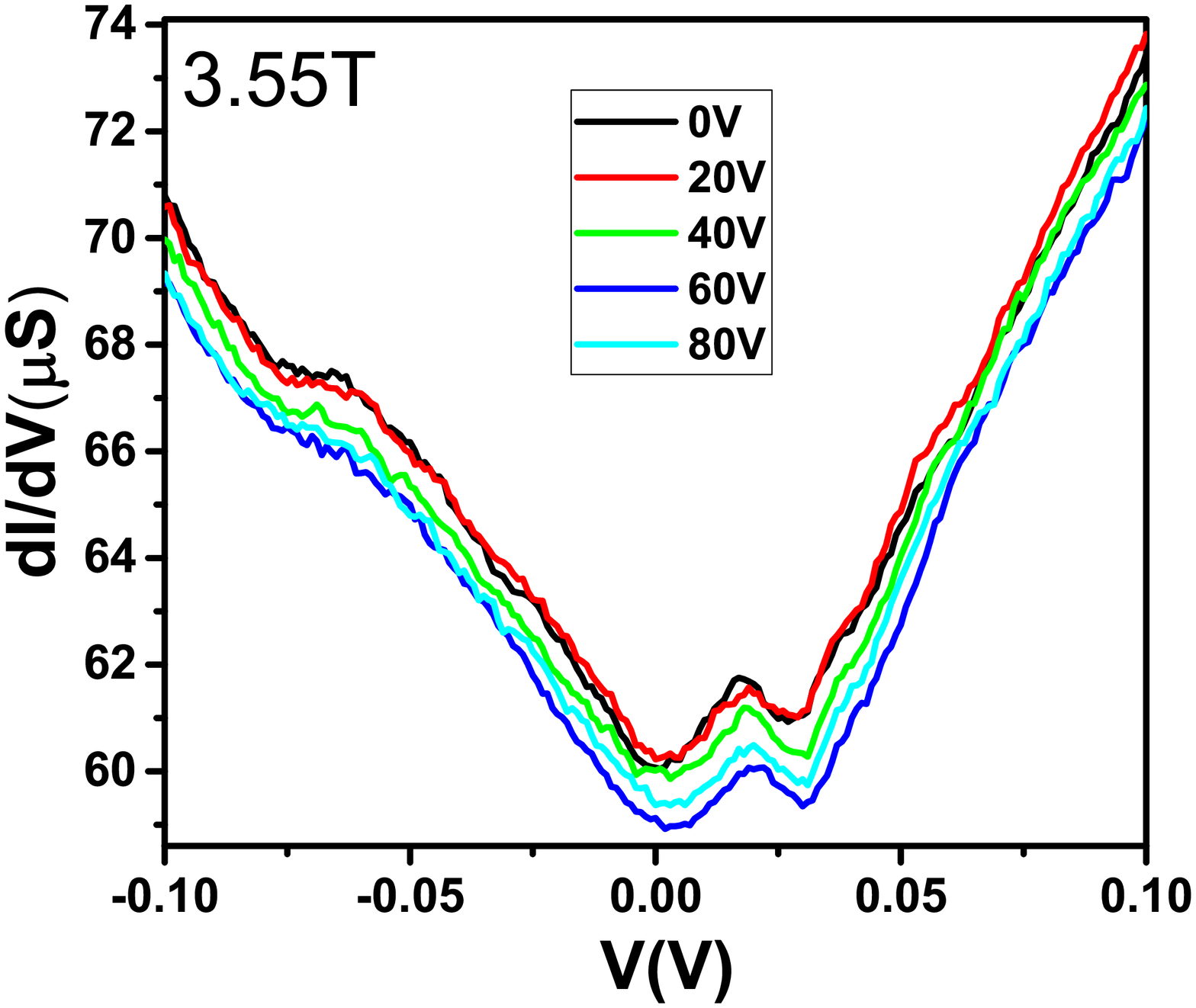} 
\hspace{-10mm}
\includegraphics[width=60mm]{dIdV_5p3T.eps}\\
\hspace{-10mm}
\includegraphics[width=60mm]{dIdV_6p2T.eps} 
\hspace{-10mm}
\includegraphics[width=60mm]{dIdV_7p65T.eps} 
\hspace{-10mm}
\includegraphics[width=60mm]{dIdV_10p25T.eps}
\caption{Differential conductance as a function of the bias voltage for different voltages on the actuator. Each panel shows data taken at a certain magnetic field (0T, 1.25T, 1.65T, 2.65T, 3.55T, 5.3T, 6.2T, 7.65T, 10.25T)}
    \label{dIdV_V_B-2}
\end{figure}

Fig. \ref{dIdV_V_B-2} presents the whole set of data of differential conductance for different voltages imposed on the actuators taken at different magnetic fields. A switching effect of the current across the sample occurs between $1.25\ T$ and $1.65\ T$.



\end{document}